\documentclass[amsfonts,nofootinbib,eqsecnum,floats,amssymb,amsmath,showpacs,pdflatex]{revtex4}
\ifx\pdfoutput\undefined
\usepackage{graphicx}
\else
\usepackage[pdftex]{graphicx}
\usepackage{epstopdf}
\fi
\newcommand\beq{\begin{equation}}
\newcommand\eeq{\end{equation}}
\newcommand\bea{\begin{eqnarray}}
\newcommand{\SR}{\left( \pm \right)}
\newcommand{\SP}{\left( + \right)}
\newcommand{\SM}{\left( - \right)}
\newcommand\eea{\end{eqnarray}}\def\ub#1{\underleftarrow{#1}}
\usepackage{amsmath,latexsym,eufrak,mathrsfs,pxfonts,bbm,parskip}
\makeatletter
\newcommand{\pback}[1]{{
   \let\@rrow=\leftarrowfill
   \mathchoice{\AIN@stemPullBack{#1}{\@rrow}}{\AIN@stemPullBack{#1}{\@rrow}}
     {\AIN@indxPullBack{#1}{\@rrow}}{\AIN@indxPullBack{#1}{\@rrow}}}
   \vphantom{#1}}

\newcommand{\AIN@stemPullBack}[2]{
   \vtop{\mathsurround=0pt
   \ialign{##\crcr$\textstyle{#1}\strut$\crcr
     \noalign{\kern-0.4ex\nointerlineskip}{\tiny#2}\crcr}}}

\newcommand{\AIN@indxPullBack}[2]{
   \vtop{\mathsurround=0pt
   \ialign{##\crcr\hfil$\scriptstyle{#1}$\hfil\crcr
     \noalign{\kern+0.4ex\nointerlineskip}{\tiny#2}\crcr}}}
\makeatother

\let\puto=\overcirc

\def\l{\ell}

\def\ba{\begin{eqnarray}}
\def\ea{\end{eqnarray}}
\def\be{\begin{equation}}
\def\ee{\end{equation}}
\def\puto#1{\rlap{\raise.5ex\hbox{\char'27}}{#1}}

\begin{document}

\title{Horizon Mechanics and Asymptotic Symmetries with a Immirzi-like Parameter in 2+1 Dimensions}

\author{Rudranil Basu}\email{rudranil@bose.res.in}
\affiliation{S. N. Bose National Centre for Basic Sciences \\ 
JD Block, Sector III, Salt Lake City, Calcutta 700098, India}
\author{Ayan Chatterjee}\email{achatterjee@imsc.res.in}
\affiliation{The Institute of Mathematical Sciences \\
CIT Campus, Taramani, Chennai 600113, India}

\begin{abstract} Starting with a generalized theory of $2+1$ gravity containing an Immirzi like parameter, we derive 
the modified laws of black hole mechanics using the formalism of weak isolated horizons. Definitions of horizon mass and 
angular momentum emerge naturally in this framework. We further go on to analyze the asymptotic symmetries, as first discussed 
by Brown and Henneaux, and analyze their implications in a completely covariant phase space framework.
\end{abstract}
\maketitle

%%%%%%%%%%%%%%%%%%%%%%%%%%%%%%%%%%%%%%%%%%%%%%%%%%%%%%%%%%%%%%%%%%%%%%%%%%%%%%%%%%%%%%%%%%
\section{\bf Introduction}
Gravity in $2+1$ dimensions \cite{djh,dj} has been an active research arena primarily 
because it is an example of exactly solvable quantum system \cite{witten_1988}. The aim of this program also lies in gaining knowledge about
 quantum gravity phenomena for the difficult problem in $3+1$ dimensions (reference \cite{carlip_book} has detail discussions). Though 
the theory is fairly understood in 
certain topological set up \cite{ezawa,basu_pal} (\emph{i.e.}, when the spatial foliations are compact Riemann surfaces), not 
everything is trivial when there is an inner boundary. Perhaps the most interesting, and hence most studied of these theories of 
$2+1$ gravity is the one with a negative cosmological constant which has been shown to admit the BTZ black hole
as an excited state and the AdS$_3$ solution as it's vacuum \cite{btz}. Although there are a more general class of black holes in 2+1 topological gravity theories \cite{Bose:1999ba}, of which BTZ is a special one. In the first order formalism, this theory is equivalent 
to $SO(2,1) \times SO(2,1)$ Chern-Simons gauge theory. Chern-Simons is again a purely topological theory. The local 
lorentz transformation and the diffeomorphisms are only local excitations, which being gauge, leave the theory devoid of any 
local physics. Global or topological degrees of freedom are the only ones to look for while constructing the physical dynamics 
of gravity in 2+1 dimensions. This is in contrast to the much studied topologically massive gravity (TMG) theory \cite{djt}
(or it's recently understood ramifications \cite{bht,liu}) in which one introduces local degrees of freedom, a parity 
violating massive graviton. It contains the usual Einstein-Hilbert term, the gravitational Chern-Simons \emph{and} a 
cosmological constant. It has also been suggested that a three dimensional gravity can always be transformed to  TMG gravity 
through field redefinition and a consistent truncation \cite{gupta_sen}. The theory of TMG has some peculiarities - the massive excitations carry 
negative energy for a positive coupling constant (in this case, it  is the $G$) \cite{djt}. In case of negative cosmological constant TMG, the 
situation is drastic. Change in sign of the coupling constant gives excitations with positive energy but gives negative mass 
BTZ black hole solutions \cite{moussa_clem}. 

Black holes in such theories (see \cite{moussa_clem,cavaglia_1999,chow_2009,egrumiller_2010} and references therein) and their
entropy have been studied in great detail \cite{kraus,hota} and in \cite{Tachikawa:2006sz} exhaustively for a large class of interactions governed by Chen Simons theory. In three dimensions, black hole entropy calculations are majorly based upon two different routes. Most popular is the one which follows \cite{strominger_1997}; this again is based on the results of the seminal 
paper by Brown and Henneaux \cite{brown_henneaux}.  They showed that asymptotic symmetries of a solution of $2+1$ general relativity 
with negative cosmological constant (not necessarily the BTZ solution), is given by a pair 
of Witt algebras- the deformation algebra of $S^1$ instead of the expected isometry $SO(2,2)$ of AdS$_3$. Canonical phase space
realization of these asymptotic symmetries however are given by a pair of Virasoro algebras, which are centrally 
extended version of the symmetry algebra. A simple use of Cardy formula for the
central extensions gives the entropy (see \cite{carlip_2005} for discussions). On the other hand, there is another path, (eg \cite{kraus, Tachikawa:2006sz}) which uses covariant phase space framework, following Wald \cite{wald}. Unfortunately this approach heavily relies on the (bifurcate) Killing horizon structures, which have their own problems including restriction to non-extremal horizons only. Dynamical issues, conserved charges in 
similar class of theories, including the canonical realization of asymptotic symmetries have been 
studied in \cite{hota,clement_2007, blagojevic_2003,blagojevic_2005,blagojevic_2008,hmt_2009, compere_2009}. Entropy of the BTZ black hole
in these modified topological theories and the TMG  were also presented in these papers. Contrary to the Bekenstein-Hawking 
expectation, the entropy turns out not only to be proportional to the black hole area, but also to 
some extra terms, involving even the horizon angular momentum \cite{hota,blagojevic_2008}. In this paper, we shall investigate related issues
for a general class of theories in a covariant manner and show that such results are expected. Moreover, we shall establish that our method is equally applicable to extremal and non-extremal black holes since it does not rely on the existence of bifurcation spheres.
   
Generalized versions of $2+1$ topological gravity which retains its topological nature came into prominence through
\cite{mielke}. In the present work, we consider a special case of such generalized theory \cite{ccgkm,bon_lev}. More precisely,
we shall work with a theory having a negative cosmological constant and a parameter which imitates the Barbero-Immirzi 
parameter of $3+1$ gravity \cite{hol}. The possibility of this generalization also was hinted in the pioneering 
work \cite{witten_1988}. The gauge group still remains $SO(2,1) \times SO(2,1)$. This theory is interestingly linked to TMG 
in a subtle way, as  discussed in \cite{basu_pal}. When one forces the torsion ($T^{I}$) to be zero, \emph{i.e.} goes 
to the partial solution space of $T^{I} =0$, one lands on TMG. The limit $ \gamma \rightarrow 1$, of the
new parameter (which behaves as chiral parameter in TMG literature), has interesting consequences both in the 
purely topological and the massive theories \cite{basu_pal,blagojevic_2005,grumiller_2010,li_song_strominger, carlip_2008, carlip_deser_2009}.

Our aim will be to establish the laws of black hole mechanics in this theory and to determine entropy in fully covariant framework, and on doing so we will show how our approach fills up the gaps in existing literature. In 
this respect, it becomes important to introduce conserved charges like the angular momentum and mass. We shall use the formalism of 
isolated horizons to address these issues. The set-up of isolated horizons is robust and conceptually straightforward, resulting 
in surprising simplicity in calculations. The 
details of isolated horizon formalism were developed in a series of papers \cite{abf_letter,afk,abl,ak_lr}. The
isolated horizon formalism for general relativity in $2+1$ dimensions was developed in \cite{adw}. We shall
however use a weaker set of boundary conditions than \cite{adw}, extend to more general theories, study the asymptotic
symmetries and eventually determine the entropy of horizons. 

 The basic 
idea is the following: a horizon (black hole or cosmological) is a null hypersurface which can be described locally, by providing 
the geometric description of that surface only. The black hole horizon (we are interested in these horizons here) is described in 
this formalism to be an internal boundary of spacetime which is expansion free and on which the field equations 
hold\footnote{This construction is more general than that of Killing horizon. Laws of black hole mechanics
 were proved for this quasi-local definition too \cite{wald, racz_wald}. But, as mentioned earlier this formalism does not seem useful to address
 extremal horizons. Improvements by introducing extremal horizons in the same 
space of the non-extremal horizons in the  isolated horizons framework were made in \cite{cg1, cg2}.}. Unlike the 
Killing horizons/event horizons, we need not look \emph{outside/asymptotic} or the space time in the vicinity of horizon
to define isolated horizons; only horizon properties are enough. It is because of this generality that 
isolated horizon is useful to describe even solutions where the asymptopia is still not well-defined or has not developed 
yet. More precisely, all event horizons/Killing horizons are isolated horizons but not all isolated horizons are 
Killing/event horizon.  As it happens (and we shall show this below), the boundary conditions enable us to prove 
the zeroth law of black hole mechanics directly. The first law of black hole mechanics and construction of 
conserved charges is not difficult in this formalism. Given a field theory, there exists a straightforward
 way which enable us to covariantly construct the space of solutions (and a symplectic structure on the covariant phase-space), initially 
introduced in \cite{abr}. This has been applied successfully to study dynamics of space times with
isolated horizon as an internal boundary. The conserved charges (like angular momentum) are
 precisely the \emph{Hamiltonian functions} corresponding to the vector field generating 
canonical transformations or the so called \emph{Hamiltonian vector fields} (which in this case is related to rotational 
Killing vector field on ). The first law is, in this description, the necessary and sufficient condition for the
 null generator of the horizon to be a Hamiltonian vector field. These features, as we shall show below, can be 
established very easily. 

There is a precise definition of the asymptotic symmetry group, if we know the fall of behaviour of the geometry asymptotically. A natural question to ask is whether the action of this group on the pre-symplectic manifold, defined through the degenerate symplectic structure, is a Hamiltonian. As have been the expectation through canonical analyses made earlier, the answer is not in the affirmative, rather the algebra of symmetry genetrators get centrally extended. This leads to finding the black hole entropy as discussed earlier (just as for the TMG case). This is similar to TMG  where the parameters are the topological mass and the 
cosmological constant. (For TMG, this implies that the massive graviton introduced through the  extra couplings have no effect on the entropy.)

Plan of the paper is as following: In the section \eqref{sec2}, we shall recall the definition of Weak Isolated Horizons (WIH) and
prove the zeroth law of black hole mechanics \cite{cg1,cg2}. The proof of zeroth law is purely kinematical and does not require any
dynamical information. In section \eqref{sec3a}, we shall first discuss the generalised theories in $2+1$ dimensions
and then introduce the theory with $\gamma$-parameter and negative cosmological constant. This section will also include 
a discussion of the BTZ solution as an example of a
black hole solution in this theory. Since we shall be interested in manifolds with inner and outer boundaries
we need to establish that the variational principle is well defined. In sections \eqref{sec3b} and \eqref{sec3c},
we shall establish that indeed the action principle is well defined even when the inner boundary is a WIH. In 
section \eqref{sec3d}, we construct the space of solutions and symplectic structure. The phase-space contains
all solutions, (extremal as well as non-extremal black hole solutions) which satisfy the boundary conditions
of WIH for the inner boundary and are asymptotically AdS at infinity.
In section \eqref{sec3e}, we shall show how the angular momentum can be extracted from the symplectic structure. The angular
momentum will naturally arise as a Hamiltonian function (on the phase-space) corresponding to the Hamiltonian vector field
associated with rotational Killing vector field on the spacetime. When the definition is applied to the BTZ solution, it will naturally arise
that the angular momentum  depends on the parameters $J$ and $M$ of the solution. In section \eqref{sec4}, we shall construct the
vector fields which generate diffeomorphisms preserving the asymptotic conditions. We shall construct Hamiltonians functions corresponding to
these vector fields and show that in presence of a WIH inner boundary, the Hamiltonian charges do not realize the algebra of
vector fields. The difference is a central extension which gives rise to the entropy for black holes in these theories. We shall
also observe that the parameter $\gamma$ shows up in all stages. We shall discuss these issues in the section \eqref{sec5}.      

%%%%%%%%%%%%%%%%%%%%%%%%%%%%%%%%%%%%%%%%%%%%%%%%%%%%%%%%%%%%%%%%%%%%%%%%%%%%%%%%%%%%%%%%%%%%%

\section{Weak Isolated Horizon: Kinematics}\label{sec2}
We now give a very brief introduction to weak isolated horizons
\cite{cg1}. Let $\mathcal M$ be a three-manifold equipped with a 
metric $g_{ab}$ of signature $(-,+,+)$. Consider a null hypersurface
$\Delta$ in $\cal M$ of which $\ell^a$ is a future directed null normal. 
However, if $\ell^a$ is a future directed null normal, so is
$\xi\ell^a$, where $\xi$ is any arbitrary positive function on $\Delta$. 
Thus, $\Delta$ naturally admits an equivalence class of null normals 
$[\,\xi\ell^a\,]$. The hypersurface $\Delta$ being null, the metric induced
on it by the spacetime metric $g_{ab}$ will be degenerate. We shall denote
this degenerate metric by $q_{ab}\triangleq g_{\ub{ab}}$ (since we are using 
abstract indices, we shall distinguish intrinsic indices on
$\Delta$ by pullback and $\triangleq$ will  mean that the equality holds
{\em only on} $\Delta$). The {\em inverse} of $q_{ab}$ will be 
denoted by $q^{ab}$ such that $q^{ab}q_{ac}q_{bd}\triangleq q_{cd}$. 
The expansion $\theta_{(\ell\,)}$ of the null normal $\ell^a$ is then defined by
$\theta_{(\ell\,)}=q^{ab}\nabla_a\ell_b$, where $\nabla_a$ is the covariant
derivative compatible with $g_{ab}$.
Null surfaces are naturally equipped with many nice properties. Firstly,
the null normal is hypersurface orthogonal and hence is twist-free. Secondly,
the $\ell^a$ is also tangent to the surface. It is tangent to the geodesics
generating $\Delta$. Thus, any $\ell^a$ in the class $[\xi\ell^a ]$
satisfies the geodesic equation:
\bea \label{kappa}
\ell^a \nabla _{\ub a} \ell ^b \triangleq \kappa _{ (\ell)}\,\ell^{b}.
\eea
We shall interpret the acceleration $\kappa _{ (\ell)}$ as the surface 
gravity. If the null normal to $\Delta$ is such that $\kappa$ vanishes,
we shall call it to be extremal surface. Otherwise, the surface will
be called non-extremal. The variation of $\kappa$ in the null class
 $ \left[ \xi \ell \right]$ being as
$ \kappa _{( \xi \ell)} = \xi \kappa _{ (\ell)} + \pounds_{\ell} \xi$.

In what follows, we shall use the Newmann-Penrose (NP) basis 
for our calculations. In three dimensions, this will consist of
two null vectors $\ell^{a}$ and $n^{a}$ and, one spacelike vector $m^{a}$
They satisfy the condition $\ell.n=-1=-m.m$ while other
scalar products vanish. This basis is particularly useful
for our set-up because the normal to
$\Delta$, denoted by $\ell^{a}$ can be chosen to be the $\ell^a$ of NP basis.
The spacelike $m^{a}$ will be taken to be tangent to $\Delta$. In this basis,
the spacetime metric will be given by $g_{ab}=-2\,\ell_{(a}n_{b)}+
2\,m_{(a}m_{b)}$ whereas the pullback metric $q_{ab}$ will be simply,
$q_{ab}\triangleq m_{a}m_{b}$.
 
 %%%%%%%%%%%%%%%%%%%%%%%%%%%%%%%%%%%%%%%%%%%%%%%%%%%%%%%%%%%%%%%%%%%%%%%%%%%%%%%%%%

\subsection{Weak Isolated Horizon and the Zeroth Law}\label{sec2a}

The null surface $\Delta$ introduced above is an arbitrary 
null surface equipped with an equivalence class of null
normals $[\xi\ell^a]$. The conditions on $\Delta$ are too 
general to make it resemble a
black hole horizon. To enrich $\Delta$ with
useful and interesting information, we need to impose
some additional structures (the imposed conditions will be weaker than that in \cite{adw}
in the sense that our equivalence class of null normals will be related by functions
on $\Delta$ rather than constants).
As we shall see, the zeroth law and the first law of black hole
mechanics will naturally follow from these conditions. These definitions
will be local and only provides a construction of black hole horizon and do not 
define a black hole spacetime which is a global object. However, if there is 
a global solution, like the BTZ one, then these conditions
will be satisfied. 

The null surface $\Delta$, equipped with an equivalence class of null
normals $[\xi\ell^a]$, will be called a {\em weak isolated horizon} (WIH)
if the following conditions hold:
\begin{enumerate}
\item $\Delta$ is topologically $S^1 \times \mathbb{R}$.
\item The expansion $\theta_{(\xi\ell)}\triangleq0$ for any $~\xi\ell^a$ in the
  equivalence class.
\item The equations of motion and energy conditions
 hold on the surface $\Delta$ and the vector field $-T^{a}_{b}\xi\ell^b$
is future directed and causal. 
\item There exists a $1$-form $\omega^{(\xi \ell)}$ such that it is 
lie-dragged along the horizon $\Delta$,
\begin{equation}\label{lielomega}
 \pounds_{\xi \ell}\,\omega  ^{(\xi \ell)} \triangleq 0
\end{equation}
\end{enumerate}

In the literature, $\Delta$ is called a \emph{non-expanding horizon} (NEH)
if it satisfies only the first three conditions. It is clear that
that the boundary conditions for a NEH
hold good for the entire class of null normals $[\xi\ell^{a}]$
if it is valid for one null normal in that class. The 
Raychaudhuri equation imply that NEHs are also 
shear free. Thus, NEHs are twist-free,
expansion-free and shear-free and this implies that the covariant
derivative of $ \ell^{a}$ on $\Delta$ much simple. There exists a
one form $ \omega ^{( \ell)}$ (see appendix \ref{New_Pen} for a
Newman-Penrose type discussion), such that
\bea \label{omega}
\nabla _{\ub a} \ell ^b \triangleq \omega _a ^{(\ell)} \ell ^b
\eea
The one form $\omega_{a} ^{(\ell)}$ varies in the equivalence class $[\xi\ell^a]$
as 
\begin{equation}\label{omegavar}
\omega^{(\xi\ell)}\triangleq\omega^{(\ell)}+d\ln\, \xi
\end{equation}
A few other conclusions also follow. Firstly, from equations
(\ref{kappa}) and (\ref{omega}), it follows that
$\kappa_{(\xi\ell)}\triangleq\xi\ell.\omega^{(\xi\ell)}$.
Secondly, that the null normals in the equivalence class are Killing vectors
on NEH $\pounds_{\ell}\, q_{ab} \triangleq 2 \ub{\nabla _{ (a} \ell _{b)}} \triangleq 0$. 
Thirdly, the volume form on $\Delta$, is lie-dragged by and null normal
in the equivalent class, $\pounds_{\xi\ell}m\triangleq0$.

 At this point one should note that the acceleration $ \kappa _{ (\xi \ell)}$ 
is in general a function on $ \Delta$. If we want NEH to obey the zeroth law 
of black hole mechanics, which requires constancy of the acceleration of
the null normal on $ \Delta$, we should restrict it further. This 
is done by demanding the fourth condition in the list of boundary conditions,
equation \eqref{lielomega}. Although this is not a single condition, 
(\emph{i.e.} unlike the other three conditions,  it is not guaranteed that
if this condition holds for a 
single vector field $ \ell^a$, it will hold for \emph{all} the others in
the class $ \left[ \xi \ell^a\right]$ for any arbitrary $\xi$), one can 
always choose a class of functions $\xi$ on 
$\Delta$ \cite{cg1, cg2}, for which this reduces to 
a single condition. For example, if the class of function is,
$\xi=F\,\mathrm{exp}(-\kappa_{(\ell)}v)+\kappa_{(\xi\ell)}/\kappa_{(\ell)}$, 
where $\ell^{a}=(\partial/\partial v)^{a}$ and $F$ is a function such that
$\pounds_{\ell}\,F\triangleq0$, the condition
holds for the entire equivalence class.\footnote{This is a virtue in disguise
in the sense that we can interpolate between extremal horizons, with 
$\kappa\triangleq 0$ to non-extremal horizons with $\kappa\ne 0$
using this $\xi$. In other words, we can use this formalism to
accommodate extremal as non-extremal horizons in the same phase space.}.
Also note that from \eqref{omegavar} that $d\omega ^{(\xi \ell)}$, which
is proportional to the Weyl tensor, is independent
of variation of $\xi$. Since the Weyl tensor vanishes identically in 
three dimensions, we have $ d \omega ^{(\xi \ell)} \triangleq 0$. The equation 
\eqref{lielomega} then gives the \emph{zeroth law}: $ d \kappa _{(\xi \ell)} \triangleq 0$.

%%%%%%%%%%%%%%%%%%%%%%%%%%%%%%%%%%%%%%%%%%%%%%%%%%%%%%%%%%%%%%%%%%%%%%%%%%%%%%%%%%%%%%%%
\section{Weak Isolated Horizon: Dynamics}\label{sec3}
In this section, we shall introduce the action for our theory and derive the
laws of black hole mechanics. We shall use the first order connection 
formulation. This formulation is tailor-made for our set-up and the 
calculational simplicity will be enormous. In particular, the construction
of the covariant phase-space and it's associated symplectic structure
is a straightforward application of the notions used
in higher dimensions \cite{cg1,cg2}. The use of forms
also simplifies the calculation of first law and the conserved charges.

Our 3-manifold $\mathcal{M}$ will be taken to be topologically 
$M\times\mathbb{R}$ with boundaries. The inner null boundary will be
denoted by $\Delta$ which is taken to be
topologically $S^{1}\times \mathbb{R}$. The initial and final space-like
boundaries are denoted by $M_{-}$ and $M_{+}$ respectively. The boundary
at infinity will be denoted by $i_{0}$. In what follows,
the inner boundary will be taken to be a WIH. In particular, this implies that
the surface $\Delta$ is equipped with an equivalent class of null-normals
$[\xi\ell^{a}]$ and follows eqn. \eqref{lielomega}.   
%%%%%%%%%%%%%%%%%%%%%%%%%%%%%%%%%%%%%%%%%%%%%%%%%%%%%%%%%%%%%%%%%%%%%%%%%%
 
\subsection{The Action in $2+1$ dimensions }\label{sec3a}
Action describing 2+1 gravity with negative cosmological constant $\Lambda = -\frac{1}{l^2}$ in first order formalism 
is (in our convention of $16 \pi G =1=c$) 
\bea \label{Palatini}
I=\int _{M}e^I \wedge \left( 2\, dA_I + 
\epsilon _I{}^{JK}\, A_J \wedge A_K \right) + \frac{1}{3l ^2}\, \epsilon ^{IJK}\, e_I \wedge e_J \wedge e_K
%I_{GR}= \frac{1}{8 \pi G} \int _{M} e^I \wedge \left(2 d \omega _I + \epsilon _I{}^{JK} \omega _J \wedge \omega _K + 
%\frac{1}{3l^2}\epsilon _I{}^{JK} e_J \wedge e_K \right)
\eea
where $e^I$ are the $SO(2,1)$ orthonormal triad frame and the $A ^I$ are connections (or canonically projected local connection) of 
the frame-bundle with structure group $SO(2,1)$. The above action is well defined and differentiable in absence of boundaries. However, as 
we will show in the next subsection, with the present boundary conditions (at internal and asymptotic) at hand, the variational problem 
is well defined with this action itself, without any boundary term. \footnote{Strictly speaking, one should add an asymptotic boundary
term to this action, which may render the whole action finite \cite{abf_letter}. But since this has nothing to do with
dynamics, \emph{i.e.} doesn't affect the variation procedure, we omit it.}

The equations of motion are expected first order versions of the Einstein equation with cosmological constant:
\bea \label{EOM1}
%\begin{subequations}
F_I := 2d A _I +  \epsilon _I{}^{JK}\, A _J \wedge A _K = -\frac{1}{l^2}\, \epsilon _I{}^{JK} \, e_J \wedge e_K \\
\label{EOM2}
T_I := d e _I + \epsilon _{IJK}\, e^J \wedge A ^K = 0
%\end{subequations}
\eea
More general models for 2+1 gravity with negative cosmological constant were introduced in
\cite{mielke} and later studied extensively in \cite{ccgkm,blagojevic_2003,blagojevic_2005}, which without matter fields read:
\bea \label{mielke}
I = a \, I_1 +b \, I_2 + \alpha _3\,  I_3 + \alpha _4 \, I_4
\eea
where
\begin{eqnarray*} 
&I_1& =\int _{M} e^I \wedge \left(2 d A _I + \epsilon _I{}^{JK} A _J \wedge A _K \right) \\
&I_2 &= \int _{M} \epsilon ^{IJK} e_I \wedge e_J \wedge e_K \\
&I_3 &= \int _{M} A^I \wedge d A _I +  \frac{1}{3} \epsilon _{IJK}\, A^I \wedge A^J \wedge A^K \\
&I_4& = \int _{M} e^I \wedge d e_I + \epsilon _{IJK}\, A ^I \wedge e^J \wedge e^K
\end{eqnarray*}
For more references on various applications of this model and its relevance with topological massive gravity (TMG) and chiral TMG, see
\cite{basu_pal} and references therein. However one must note that this model does not reproduce the Einstein 
equations  \eqref{EOM1} and \eqref{EOM2} for arbitrary values of the parameters $ a,b, \alpha _3, \alpha _4$. We choose, as 
a special case of the above model, those values of these parameters which gives the expected equations of motions 
\bea \label{para_choice}
a=1 ~~ b = \frac{1}{3  l^2} ~~ \alpha _3 = \frac{l}{\gamma} ~~ \alpha _4 = \frac{1}{ \gamma l}
\eea
$\gamma $ is introduced as new dimensionless parameter from 2+1 gravity perspective. Effectively
 \eqref{para_choice} is the equation of a 3 dimensional hypersurface parametrized by $G,l, \gamma$ in the 4-d parameter 
space of $a,b, \alpha _3 , \alpha _4$.
\\
In \cite{witten_1988}, it was established that first order 2+1 gravity can be written as Chern Simons gauge theory, the 
gauge group being determined by the sign of the cosmological constant. It is now explicit that 2+1 gravity does not have any local
 degrees of freedom, since Chern Simons theory is a topological one. Moreover, as a unique feature of 3 dimensions, all the 
invariances of first order gravity, ie the local Lorentz ($SO(2,1)$) transformations and arbitrarily large number of diffeomorphisms 
are now taken care of by finite dimensional Chern Simons gauge group, when viewed on shell. 
Following \cite{witten_1988, ccgkm},  one can introduce the $SO(2,1)$ or equivalently $SL(2, \mathbbm{R})$ or 
$SU(1,1)$ connections for a principal bundle over the same base space of the frame bundle:
\begin{eqnarray*}
\mathcal{A}^{\SR} := \left( A ^I \pm \frac{e^I}{l} \right) J_I^{\SR}~.
\end{eqnarray*}
Now, it also happens that $J_I^{\SR}$ form two decoupled $SO(2,1)$ lie algebras:
\begin{eqnarray}
\left[J^{\SP}_I , J^{\SP}_J\right] &=& \epsilon _{IJK}\, J^{\SP K} \quad  \left[J^{\SM}_I , J^{\SM}_J\right] = \epsilon _{IJK} \,J^{\SM K} \nonumber \\
&& \left[J^{\SP}_I , J^{\SM}_J\right] = 0.
\end{eqnarray}
The metric on the Lie algebra is: 
$$\langle J^{\SR I}, J^{\SR J} \rangle = \frac{1}{2} \eta ^{IJ}$$
It is easily verifiable that the action 
\bea \label{split}
\tilde I=l \left( I^{\SP} - I^{\SM}\right)
\eea
is same as \eqref{Palatini} upto boundary terms which are guaranteed to vanish in our case. 
Where
\bea \label{defn_I}
I^{\SR} = \int _{M}\mathrm{tr}\left( \mathcal{A}^{\SR } \wedge d \mathcal{A}^{\SR} + \frac{2}{3}  \mathcal{A}^{\SR} \wedge \mathcal{A}^{\SR} \wedge 
\mathcal{A}^{\SR } \right)
\eea

One striking feature of this formulation is that the last two terms of \eqref{mielke} can also be incorporated in terms 
of $ \mathcal{A} ^{\SR}$, for ($ \alpha _3 = l^2 \alpha _4$) as:
$$I^{\SP} + I^{\SM} = \int _{M}\left( A^I \wedge d  A _I + \frac{1}{l^2} e^I \wedge d e_I + 
\frac{1}{3} \epsilon _{IJK}  A^I \wedge A^J \wedge A^K + \frac{1}{l^2}\epsilon _{IJK} A^I \wedge e^J \wedge e^K \right)$$
and the same equations of motion \eqref{EOM1} and \eqref{EOM2} are also found from varying this action. In this paper, we 
shall work with the following action:
\bea 
\label{new}
I &=&l \left( I^{\SP} - I^{\SM}\right) + \frac{l}{\gamma} \left( I^{\SP} + I^{\SM}\right) 
\nonumber\\
&=& l \left[ \left(1/\gamma +1 \right) I^{\SP} + \left(1/\gamma -1 \right) I^{\SM}\right]
\eea
with a dimensionless non-zero coupling $\gamma$. This action \eqref{new} upon variations with respect to
 $\mathcal{A}^{\SP}$ and $\mathcal{A}^{\SM}$ give equations of motion as expected from Chern-Simons theories. This 
imply that the connections $\mathcal{A}^{\SR}$ are flat:
\bea \label{flat}
\mathcal{F}^{\SR}_I : = d\mathcal{A}^{\SR}_I + \epsilon_{IJK} \mathcal{A}^{\SR J} \wedge \mathcal{A}^{\SR K} = 0.
\eea 
It is also easy to check that the above flatness conditions of these $SO(2,1)$ bundles \eqref{flat} are equivalent to
 the equations of motion of general relativity \eqref{EOM1}, \eqref{EOM2}. 
Notice that the new action is like the Holst action \cite{hol}  used in 3+1 gravity. In our case the 
parameter $\gamma$ can be thought of being the 2+1 dimensional counterpart of the original Barbero-Immirzi parameter. Moreover the 
part $[I^{\SP} + I^{\SM}]$ of the action in this light qualifies to be at par with the topological (non-dynamical) term one adds
 with the usual Hilbert-Palatini action in $3+1$ dimensions, since this term we added (being equal to a Chern-Simons action for
 space-times we consider) is also non-dynamical. But more importantly the contrast is in the fact that the original action, which is
 dynamical in the 3+1 case is also non-dynamical here, when one considers local degrees of freedom only. However, there is a difference 
between the original B-I parameter and the present one. In the $3+1$ dimensions, 
$\gamma$ parameterizes canonical transformations in the phase space of general relativity. From the 
canonical pair of the $SU(2)$ triad (time gauge fixed and on a spatial slice) and spin-connection one goes on 
finding an infinitely large set of pairs parameterized by $\gamma$. The connection is actually affected by this 
canonical transformation, and this whole set of parameterized connections is popularly known as the
 Barbero-Immirzi connection. The fact that this parameter induces canonical transformation can be checked by seeing that the 
symplectic structure remains invariant under the transformation on-shell. On the other hand for the case at hand, \emph{i.e.} $2+1$ gravity, as
 we will see in the following sub-section that inclusion of finite $\gamma$ is not a canonical transformation.

Just like in $2+1$ gravity with a negative cosmological constant, the $2$ parameter family of BTZ black holes is a 
solution of this theory. In the standard coordinates, the solution is given by:
\begin{equation}\label{btz_sol}
ds^{2}=-N^{2}\,dt^{2}+N^{-2}\,dr^{2}+r^{2}(N^{\phi}\,dt+d\phi)^{2},
\end{equation}
where the lapse and the shift variables contain the two parameters $M$ and $J$ and are defined by:
\begin{equation}
 N^{2}=(-\frac{M}{\pi}+\frac{r^{2}}{l^{2}}+\frac{J^{2}}{4\pi r^{2}})~~~~~\mbox{and}~~~~N^{\phi}=-\frac{J}{2\pi r^{2}}
\end{equation}
The horizon is defined through the zeros of the lapse function $N^{\phi}$ which gives the position of the horizon to be:
\begin{equation}
r_{\mp}=l\,\left[\frac{M}{2\pi}\left\{1\mp(1-({J}/{Ml})^{2})^\frac{1}{2}\right\}\right]^{\frac12} 
\end{equation}
It is not difficult to see that the outer horizon (at $r_{+}$) satisfies the conditions of WIH $\Delta$. It is a 
null surface with null normal $\ell^{a}=(\partial/\partial v)^{a}+N^{\phi}(r_{+})\,(\partial/\partial \phi)^{a}$. A simple calculation 
also shows that $\theta_{(\ell)}\triangleq0$. In what follows, we shall always refer back to this solution to check if our 
definitions for conserved charges are consistent.
%\textbf{Please write about 3 dimensional action in general,
%2 lines about Witten's work,
%then some explanation about the choice of  gauge group, and the difference 
%with our work where a different inner product is chosen in the $\gamma$ 
%dependent term. Then 2 lines about other works with references.} 

%We would describe the theory describing 2+1 gravity with negative 
%cosmological constant by the following Lagrangian three form $\mathbb{L}$:

%\begin{eqnarray} \label{action}
%\mathbb{L}\, &=&  \frac{1}{16 \pi G} \Bigg[e^I \wedge \left( 2\, dA_I + 
%\epsilon _I{}^{JK}\, A_J \wedge A_K \right)  + 
%\frac{l}{ \gamma}  \left(A^I \wedge d A_I +
%\frac{1}{3}\epsilon_I{}^{JK}\, A^I \wedge A_J \wedge A_K \right) \nonumber\\&&
%+ \frac{1}{l \gamma} \left(e^I \wedge d e_I + \epsilon _{IJK}\, e^I \wedge A^J 
%\wedge e^K \right)\Bigg]  + 
%\frac{1}{3l ^2}\, \epsilon ^{IJK}\, e_I \wedge e_J \wedge e_K
%\end{eqnarray}
%\textbf{ PLEASE PUT THESE LINES BEFORE THE ACTION This theory for describing 2+1 gravity was first mentioned by \cite{Witten:1988hc} and  Mielke {\it et al} \cite{Mielke:1991nn},\cite{Baekler:1992ab}; it was later studied extensively in \cite{Cacciatori:2005wz}. For more references on various applications of this model and its relevance with topologically massive gravity, see \cite{Basu:2009dy} and references within it.}
%%%%%%%%%%%%%%%%%%%%%%%%%%%%%%%%%%%%%%%%%%%%%%%%%%%%%%%%%%%%%%%%%%%%%%%%%%%%
\subsection{Computing tetrads and connection on $\Delta$}\label{sec3b}
Before proceeding with the variation of the action and determining the 
equations of motion, it will be useful to have the values of the tetrad
and connection on the null surface $\Delta$. The usefulness of such 
calculation  will be apparent soon. We shall assume that 
it is possible to fix an internal null triad
$(\ell^{I}, n^{I}, m^{I})$ such that $\ell^{I} n_{I}=-1=- m^{I} m_{I}$ 
and all others zero. The internal indices will be raised and 
lowered with $\eta_{IJ}$. Given the internal triad basis 
$(\ell^{I}, n^{I}, m^{I})$  and $e^{I}_{a}$, the spacetime null basis 
 $(\ell^{a}, n_{a}, m^{a})$ can be constructed. We shall further assume that
the internal basis is annihilated by the partial derivative operator,
$\partial_{a}\,(\ell^{I}, n^{I}, m^{I})=0$.

Using the expression of the spacetime metric in NP basis and the internal
metric, we can write the tetrad $e^{I}_{a}$ on WIH $\Delta$ as:
\begin{equation}\label{tetradexp}
e_{\ub a}^{I}\triangleq -n_{a}\ell^{I}+m_{a}m^{I}
\end{equation} 

To calculate the expression of connection on $\Delta$, we shall use
the NP coefficients which can be seen in the covariant derivatives
of the NP basis. They are as follows:
\begin{eqnarray}\label{expcovderbasis}
\nabla_{\ub{a}}\ell^{b}&\triangleq &\omega^{(\ell)}_{a}\, \ell^{b}\\
\nabla_{\ub{a}}n_{b}&\triangleq & -\omega^{(\ell)}_{a}\, n_{b}
+U^{(\ell,m)}_{a}m_{b}\\
\nabla_{\ub{a}}m^{b}&\triangleq & U^{(\ell,m)}_{a}\, \ell^{b},
\end{eqnarray}
where, the superscripts on the one-forms $\omega^{(\ell)}_{a}$ and 
$U^{(\ell,m)}_{a}$ indicate that they depend on the transformations
of the corresponding basis vectors. The one-forms used in the
eqn. \eqref{expcovderbasis} are compact expression of the NP coefficients.
They are given by:
\begin{eqnarray}\label{expcovderbasis1}
\omega^{(\ell)}_{a}&\triangleq & (-\epsilon\, n_{a}+\alpha\, m_{a})\\
U^{(\ell,m)}_{a} &\triangleq & (-\pi\,n_{a}+\mu\, m_{a})
\end{eqnarray}

We will now demonstrate how the Newmann-Penrose coefficient $\alpha$ is fixed to be real number on $\Delta$ using 
topological arguments. Note from previous discussion that $ d \,\omega^{(\ell)} \triangleq 0$. From the 
definition \eqref{nablam} we have $ \pback{dm} \triangleq - \rho m \wedge n$. But because $\Delta$ is expansion-free and $\ell^a$ is
the generator of $\Delta$, $ \rho \triangleq 0$. Hence $m^a$ is also 
closed on $ \Delta$ ($m$ should not strictly be exact since $ \int _{S_{\Delta}} m \sim $ area of horizon $ \neq 0$). Since 
the first cohomology group of $ \Delta \simeq \mathbb{R} \times S^1 \equiv \mathbb R$ is non-trivial, we have 
in general neither $ \omega^{(\ell)}$ nor $m_{a}$ exact. Hence there exists smooth function $\varsigma$  and a real number $s$ for which 
\begin{eqnarray}\label{cohom}
\omega^{(\ell)} \triangleq d\,\varsigma + s\,m
\end{eqnarray} 
We now introduce a potential $ \psi _{(\ell)}$ for surface gravity (or the acceleration for $ \ell^a$)
$ \kappa _{(\ell)} \triangleq \ell ^a \omega^{(\ell)} _a\triangleq \epsilon$ through
$$ \pounds_{\ell} \,\psi _{(\ell)} \triangleq \kappa_{(\ell)} .$$
Since the zeroth law implies constancy of $ \kappa _{(\ell)}$ on $ \Delta$, $ \psi_{(\ell)}$ can only 
be function of $v$ (could be treated as the affine parameter on $\Delta$) only. Hence 
$\pounds_m \psi _{(\ell)} \triangleq 0$, which implies on the other hand 
$d\,\psi _{(\ell)} \triangleq -\epsilon\, n $ and $\omega\triangleq d \,\psi_{(\ell)} + \alpha\, m$. It is tempting to 
choose $ \varsigma = \psi_{(\ell)} $ by compared with \eqref{cohom}. That could only be supported 
if $ \Delta$ is axisymmetric. (Because even after choosing a triad set for which $ \underleftarrow dn \triangleq 0$, we end 
up with $\underleftarrow {d \alpha} \wedge m \triangleq 0$, which renders $ \underleftarrow {d \alpha}\triangleq 0$ only 
if $ \alpha$ is axisymmetric). For that case, we conclude $\omega ^{(\ell)}\triangleq 
(\,d \,\psi ^{(\ell)} + \alpha\,m\,)$, $ \alpha \in \mathbb{R}$. 

Now, to calculate the connection, we use two facts. First is that the tetrad is
annihilated by the covariant derivative, $\nabla_{a}\,e^{I}_{b}=0$ and,
secondly that partial derivative annihilates the NP internal basis so that
\begin{equation}\label{connectiondef}
\nabla_{\ub{a}}\ell^{I}\triangleq A_{a}^{IJ}\,\ell_{J}.
\end{equation}

Using equations \eqref{expcovderbasis1} and \eqref{connectiondef}
and $\epsilon_{IJK}=3!\,\ell_{[I}n_{J}m_{K]}$, we get the following expression
for pulled-back connection on $\Delta$:
\begin{equation}\label{connexpDelta}
A^{I}_{\ub{a}}\triangleq - U^{(\ell,m)}_{a}\,\ell^{I}+\omega^{(\ell)}_{a}\, m^{I}.
\end{equation}
The equation \eqref{connexpDelta} will be used frequently in what follows.

\subsection{Variation of the action}\label{sec3c}
\begin{figure}[t]
\begin{center}
\includegraphics[scale=0.5]{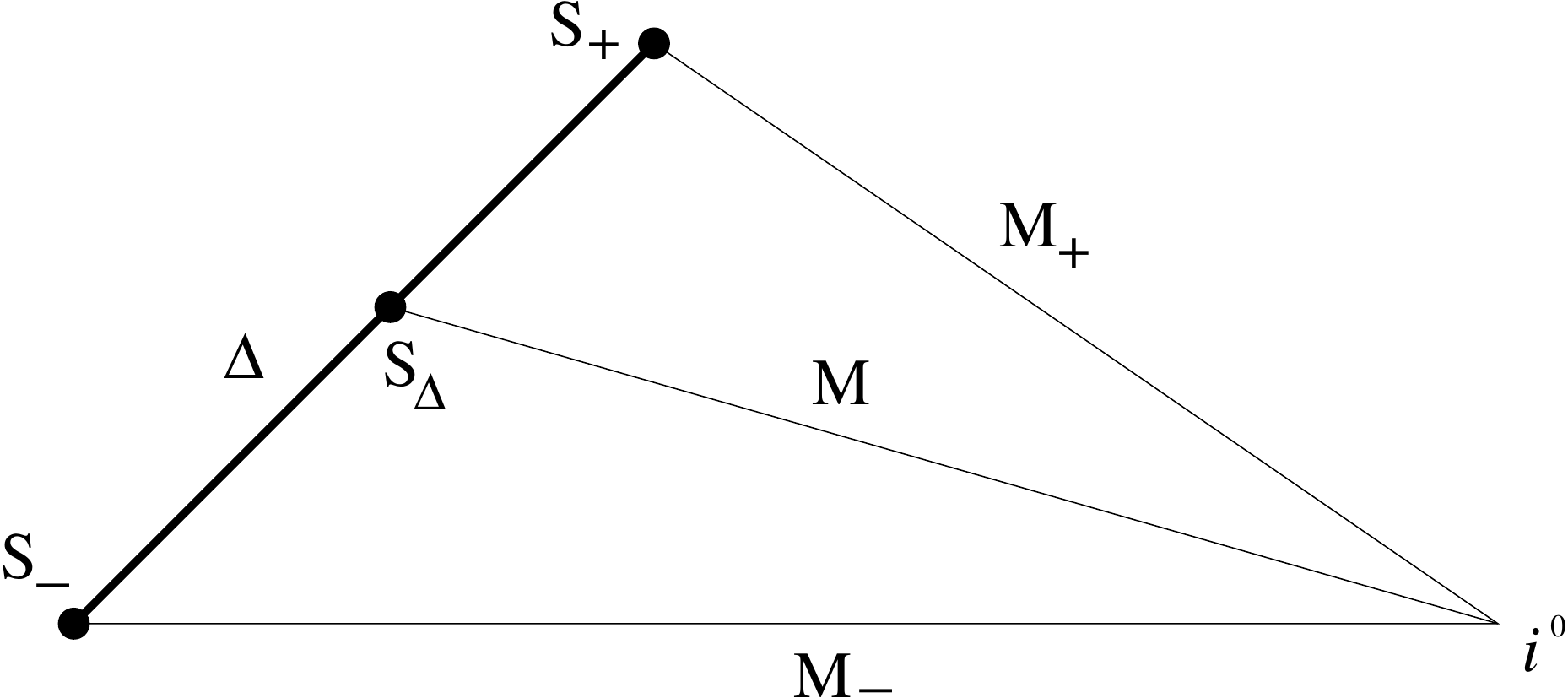}
\end{center}
\caption{Description of spacetime used in the paper. The spacetime is bounded by $2$-dimensional surfaces $\Delta, M_{\mp}$ and the infinity.
The horizon $\Delta$ is a $2$-dimensional null surface and $M_{\mp}$ are initial and final hypersurfaces. The infinity is AdS if we work
with a spacetime with negative cosmological constant.}\label{fig}
\end{figure}
In the subsection \eqref{sec3a}, we demanded that we get the first order Einstein equations of motion even
 by varying the generalized action. For spacetime manifolds without boundary, this is trivial to check. The task is now 
to vary the action to obtain the equations of motion
and also to verify that the action principle is obeyed in presence
of the boundaries. The variation will be over configurations
which satisfy some conditions at infinity and at the inner boundary (see fig. \eqref{fig}).
At infinity, they satisfy some asymptotic conditions which
are collected in the appendix of \cite{adw}. On the inner boundary $\Delta$,
they are subjected to the following conditions: $(a)$ the tetrad ($e$)
are such that the vector field $\ell^{a}=e^{a}_{I}\ell^{I}$ belongs
to the equivalence class $[\xi\ell^{a}]$ and $(b)$ $\Delta$ is a WIH.
On variation, we shall get equations of motion and some surface terms.
The surface terms at infinity
vanish because of the asymptotic conditions whereas, as we shall show,
those at WIH also vanish because of WIH boundary conditions.

Variation of  the action with respect to the tetrad ($e$) and 
connection ($A$) leads to (for $\gamma^{2}\ne 1$):
\begin{eqnarray} \label{eom}
 de^I + \epsilon ^I{}_{JK}\, e^J \wedge\, A^K &=& 0  \\
\mbox{and} ~~~~~~~~ dA _I +\frac{1}{2}\, \epsilon _I{}^{JK} \, A_J \wedge\, A_K 
&=& - \frac{1}{2l^2}\,\epsilon ^{IJK} \,e_I \wedge e_J \wedge e_K .
\end{eqnarray}
The first equation above just points out that the connection $A _I $ is a
spin-connection and the second equation is the Einstein equation.
Let us now concentrate on the surface terms. The terms on the
initial and final hyper surfaces $M_{-}$ and $M_{+}$ vanish
because of action principle. Those at the asymptotic boundary vanish because
of the fall-offs at infinity. On $\Delta$, these are
given by:
\begin{eqnarray}
\delta\, I=- \int_{\Delta}(2\,m\wedge \delta\,\omega^{(\ell)}
+\frac{l}{\gamma}\,\omega^{(\ell)}\,\wedge\,\delta\,\omega^{(\ell)}+ 
\frac{1}{l\gamma}m\,\wedge\,\delta\, m)
\end{eqnarray}

Our strategy will be to show that the integral is constant
on $\Delta$ and the integrand is a total derivative
so that the integral goes on to the initial and the final boundaries
where the variations are zero by assumption. This will then imply that
the integral itself vanishes on $\Delta$. Note that in
the above equation, $ \delta\,\omega^{(\ell)}$ refers to
the variation in $\omega^{(\ell)}$ among the configurations
in the equivalence class $[\xi\ell^{a}]$. The relation between these
are precisely given by eqn. \eqref{omegavar}. Now, we consider the
lie derivative of the integrands by ${\xi\ell}$. Since $dm\triangleq 0$,
it follows that $\pounds_{\xi\ell}\,m \triangleq 0$ 
and $\pounds_{\xi\ell}\,\omega^{(\ell)}\triangleq d(\pounds_{\xi\ell}\,ln \xi)$.
Thus, in the first term, the total contribution is on the 
initial and final hyper surfaces $M_{-}$ and $M_{+}$ where the variations
vanish. Identical arguments for the second and the
third integrands also show that the corresponding integral vanishes.
Thus, the integral is lie dragged on $\Delta$ and since the 
variations are fixed on the initial and final hyper surfaces, 
the entire integral vanishes and the action principle remains well-defined.

%%%%%%%%%%%%%%%%%%%%%%%%%%%%%%%%%%%%%%%%%%%%%%%%%%%%%%%%%%%
\subsection{Covariant Phase Space}\label{sec3d}
Analysis of the dynamics of this theory has been considerably worked out in literature \cite{blagojevic_2003} in the canonical 
framework even in presence of asymptotic boundary. A covariant phase space \cite{abr} analysis for the same theory 
was presented in \cite{basu_pal} although in absence of boundaries. As we progress, we will see how apt the
 covariant analysis is in understanding horizon phenomena and even the conserved charges arising from asymptotic 
symmetries; using the general ideas of symplectic geometry. The covariant phase space is by definition the space of 
classical solutions \eqref{eom} which satisfy the boundary conditions specified in the previous subsections.  In other words, the 
covariant phase space $\Gamma$ will consist of of solutions of the field equations which satisfy the boundary 
conditions of WIH at $\Delta$ and have fall-off conditions compatible with asymptotic conditions. In order to equip this 
space with a symplectic structure \footnote{To be more precise, here we will be dealing with  the pre-symplectic structure, since
 the theory has gauge redundancy, which appear as 'degenerate directions' for the symplectic 2-form}, we find the symplectic 
potential from variation of the Lagrangian:
\bea \label{theta}
\delta\, \mathbb{L} = d\, \Theta ( \delta ) + \mbox{terms vanishing on shell},
\eea
For the Lagrangian in hand given by \eqref{new}, the symplectic potential is given by:
\begin{equation}\label{symppot}
\Theta(\delta)=-2\, (e^{I}\,\wedge \delta\,A_{I})-
\frac{l}{\gamma}\, (A^{I}\,\wedge \delta\,A_{I})
-\frac{1}{l\gamma}\,(e^{I}\wedge \delta\,e_{I}).
\end{equation}
Upon antisymmetrized second variation, it gives the symplectic current $J$
which is a phase-space two-form. For two arbitrary vector fields 
$\delta_{1}$ and $\delta_{2}$ tangent to the space of solutions, the symplectic current for \eqref{symppot}
is given by following closed two form: 
\bea \label{J}
J( \delta _1, \delta _2) &=& \delta _1\, \Theta (\delta _2) - 
\delta _2 \,\Theta (\delta _1) \nonumber\\
&=& - 2\left[\left( \delta _1\, e^I \wedge \delta _2\, A_I- 
\delta _2\, e^I \wedge \delta _1\, A_I \right) + 
\frac{l}{\gamma}\, \delta _{1}\, A ^I \wedge \delta _{2}\, A_I + 
\frac{1}{\gamma l}\, \delta _{1}\, e^I \wedge \delta _{2}\, e_I \right]
\eea
Since the symplectic current is closed, $d\,J( \delta _1, \delta _2)=0$,
we define the presymplectic structure on the phase-space by:
\begin{equation}
\Omega\left( \delta _1, \delta _2\right)=\int_{M_{1}\cup M_{2}\cup
\Delta\cup i_{0}}\,J\,( \delta _1, \delta _2),
\end{equation}
where the terms under the integral show contributions from the
various  boundaries (refer to figure \eqref{fig}). The surfaces $M_{1}$ and $M_{2}$ are 
partial Cauchy slices inside the spacetime which meet $\Delta$ in $S_{1}$ and $S_{2}$ respectively. To show that the symplectic structure
is independent of the choice of Cauchy surface, we again consider the function 
$\psi_{(\ell)}$ such that $\pounds_{\ell}\,\psi_{(\ell)}=\kappa_{(\ell)}$ and $\psi _{(\ell)}$
vanishes on $S^{1}$ (where the affine parameter $v=0$). Choosing a orientation, it is not difficult to show that
$J\,( \delta _1, \delta _2)\triangleq dj\,( \delta _1, \delta _2)$ so that
\begin{equation}
(\int_{M_{1}}-\int_{M_{2}})\,J\,( \delta _1, \delta _2)=
(\int_{S_{1}}-\int_{S_{2}})\,j\,( \delta _1, \delta _2)
\end{equation}
which establishes the independence of symplectic structure on choice of Cauchy surfaces.The pre-symplectic 
structure on the space of solutions of the theory in presence of $ \Delta$ turns out to be
\bea \label{symp}
\Omega \left( \delta _1, \delta _2\right)&=& -2\int _M 
\left[\left( \delta _1 e^I \wedge \delta _2 A_I-
 \delta _2 e^I \wedge \delta _1 A_I \right) + \frac{l}{\gamma}\, 
\delta _{1} A ^I \wedge \delta _{2} A_I + \frac{1}{\gamma l}\,
 \delta _{1} e^I \wedge \delta _{2} e_I \right] \nonumber\\
&-& 2\int _{S^1} \left(\delta _{1} \psi_{(\ell)}\, \delta _{2}\left[\left(\frac{l \alpha}{ \gamma}+1\right) m\right] - 
\delta _{2} \psi_{(\ell)}\, \delta _{1}\left[\left(\frac{l \alpha}{ \gamma}+1\right) m\right] \right)
\eea
We shall use \eqref{symp} to define conserved quantities like the angular momentum and prove the first law in 
the next two subsections. We shall also construct the algebra of conserved charges using this symplectic structure and
obtain the entropy for black holes in this theory.
%%%%%%%%%%%%%%%%%%%%%%%%%%%%%%%%%%%%%%%%%%%%%%%%%%%%%%%%% 
\subsection{Angular Momentum}\label{sec3da}
We shall first introduce the concept of angular momentum starting from the symplectic structure, equation \eqref{symp}. Let
us consider a fixed vector field $\varphi^{a}$ on $\Delta$ and all those spacetimes which will
have $\varphi^a$ as the rotational Killing vector field on $\Delta$.  The field $\varphi^{a}$ is assumed to satisfy certain properties.
First, it should lie drag all fields in the equivalence class $[\xi\ell^a]$ ans secondly, it has closed orbits and affine 
parameter $ \in [0,2 \pi)$. To be more precise, we can construct a submanifold $\Gamma_{\varphi}$ of the covariant phase space $\Gamma$
the points of which are solutions of field equations which admit a WIH $(\Delta, [\xi\ell^a], \varphi^{a})$ with a rotational 
Killing vector field $\varphi^a$ such that $ \pounds_{ \varphi} q_{ab} \triangleq0 $,
$\pounds_{ \varphi} \omega ^{(l)} \triangleq 0$. Now, let us 
choose a vector field $ \phi$ in $ \mathcal{M}$ for each point in $\Gamma_{\varphi}$ such that
it matches with $\varphi^a$ on $\Delta$.

We shall now look for phase space realization of diffeomorphisms
generated by this vector field  $\phi^a$ on spacetime. Corresponding to the diffeomorphisms on spacetime, we can 
associate a motion in the phase space $\Gamma_{\varphi}$ which is generated by the vector field $\delta_{\varphi}=\pounds_{\varphi}$.
It is expected that the vector field $\delta_{\phi}$ will be Hamiltonian (\emph{i.e.} generate canonical transformations). In that
case, the Hamiltonian charge for the corresponding to the rotational Killing vector field can be called the angular momentum
\footnote{Since the theory we started with is background independent (has manifest diffeomorphism invariance in bulk) it is
 natural to expect that Hamiltonians generated by space time diffeomorphisms must consist of boundary terms, if any.}. In short,
this implies that $\Omega (\delta, \delta _{ \phi}) = \delta J^{ \left( \phi\right)}$
and the angular momentum is $J^{ \left( \phi\right)}$ is given by:
\begin{eqnarray}
J^{ \left( \phi\right)} &=&  -\oint _{ S_{ \Delta}} \left[ (\varphi \cdot \omega) m + \frac{l}{2 \gamma}(\varphi \cdot \omega) \omega +
 \frac{1}{2 \gamma l} (\varphi \cdot m) m\right] + \oint _{ S_{ \infty}} \left[ (\phi \cdot A^I) e_I + 
\frac{l}{ 2\gamma}(\phi \cdot A^I) A_I+\frac{1}{ 2l\gamma}(\phi \cdot e^I) e_I\right]\nonumber\\
&=& -J_{\Delta}+J_{\infty}
\end{eqnarray}
It is then natural to interpret $J_{\Delta}$ to be the angular momentum
on $\Delta$. It is simple to check that for BTZ space-time the expressions for $J _{ \Delta}$ and $J_{ \infty}$. It follows that
$ J _{ \Delta} = (\,J - Ml/ \gamma\,)= J_{ \infty}$, leaving $ J^{ \left( \phi\right)} =0$ (Note that for $\gamma\rightarrow\infty$, we
get the value of angular momentum of BTZ black hole for GR in $2+1$ dimensions with a negative cosmological constant). That
$ J_{ \Delta} = J_{ \infty}$ is also supported by the fact that $ \phi^{a}$ is global Killing vector in BTZ solution.
However, if there are electromagnetic fields, the result differs. The value of the angular momentum at infinity $J_{ \infty}$
also gets contribution from the electromagnetic fields and $J^{ \left( \phi\right)}\neq 0$ \cite{adw}.

%%%%%%%%%%%%%%%%%%%%%%%%%%%%%%%%%%%%%%%%%%%%%%%%%%%%%%%%%%%%%%%%%%%%%%%%%%%%%%%%%
\subsection{First Law}\label{sec3e}
First law is associated with energy which implies that we should look first for a timelike Killing vector field on spacetime.
Let us consider a time-like vector field $t ^a$ in $ \mathcal{M}$ associated to each point of the phase
space (live) which gives the asymptotic time translation symmetry at infinity and becomes 
$ t^a \triangleq \xi\, \ell ^a - \Omega _{(t)} \phi ^ax$ on $ \Delta$, where $ \Omega _{(t)}$ is a constant on $ \Delta$ 
but may well vary on the space of histories. Just like in the previous subsection, we ask if the associated vector field $\delta_t$
on the phase-space $\Gamma_{\phi}$ is a Hamiltonian vector field. The associated function shall be related to the energy.
In checking so, we have:
$$  \Omega \,( \delta , \delta _t)=X^{(t)} ( \delta)\, ,$$
where,
\begin{eqnarray} \label{fl1}
X^{(t)} ( \delta) = -2\kappa _{(t)} \,\delta \left( (1+ \frac{l \,\alpha}{ \gamma} )\,a_{\Delta}\right) - 
2\Omega _{(t)}\, \delta J_{ \Delta} + X^{(t)} _{ \infty}( \delta) 
\end{eqnarray}
and $ \kappa _{(t)}$ actually the surface gravity associated with the vector field 
$ \xi\ell^{a}$. $X^{(t)} _{ \infty}( \delta)$ involves integrals of fields at asymptotic infinity and
can be evaluated using asymptotic conditions on the BTZ solution for example. A simple calculation gives:
$$ X^{(t)} _{ \infty}( \delta) = \delta \left( M - \dfrac{J}{ \gamma l}\right)$$

Now the evolution along $ t^a$ is Hamiltonian only if right hand side of \eqref{fl1} is exact on phase space. This implies
if the surface gravity is a function of area only and $\Omega _{(t)}$ a function
of angular momentum only, there exists a phase space function $ E^t _{ \Delta}$ such that the \emph{first law} appears:
\begin{eqnarray}\label{1stlaw}
\delta E^t _{ \Delta} &=& \left[\kappa _{(\xi\ell)} \,\delta \left( \left(1+ l \alpha/ \gamma \right) a_{\Delta}\right) + \Omega _{(t)} \,
\delta J_{ \Delta}\right] \nonumber\\
                                &=& \left(\kappa _{(\xi\ell)} \,\delta \tilde a_{\Delta} + \Omega _{(t)} \,\delta J_{ \Delta}\right)
\end{eqnarray}
where $ \tilde{a}_{ \Delta } =  \left(1+ \dfrac{l \alpha}{ \gamma} \right) a_{\Delta}$. The presence
of $\kappa_{(\xi\ell)}$ in the first law indicates that the first law is same for both extremal
and non-extremal black holes. A mere choice of the function $\xi$ can help us interpolate between these class of solutions. We note here 
that modification in the symplectic structure of the theory leaves its footprint through $\gamma$ in the first law of (weak) isolated horizon 
mechanics. The term that plays the role of the `area' term as it appears in this first law differs 
from the standard geometrical area of the horizon. If we restrict ourselves to the class of BTZ horizons,
 we have, \footnote{In our conventions, the double roots $r_+, r_-$ of the BTZ lapse polynomial are related 
with BTZ ( $\gamma \mapsto \infty$) mass ($M$) and angular momentum ($J$) as $$ M = 2\pi \,\frac{r_+^2 + r_-^2}{l^2} ~~\mbox{and} ~~J = 4\pi\,\frac{r_+ r_-}{l}$$} 
%$\alpha \triangleq - \dfrac{ r_{ -}}{ l r_{+}} = -\dfrac{\pi J}{ a _{\Delta}^2}$.
\begin{eqnarray}\label{mod_area}\tilde{a}_{\Delta} = 2\pi \left( r_+ - r_-/{\gamma}\right)= a_{\Delta}- \dfrac{l\pi J}{\gamma a_{\Delta}}\end{eqnarray}
\subsection{Admissible Vector Fields and Horizon Mass}
In the previous discussion we used the Hamiltonian evolution of the live time vector field $t^a$ to deduce the first law. It 
is necessary and sufficient for the existence of the Hamiltonian function $E _{\Delta}^t$ as in \eqref{1stlaw} that the 
functions $ \kappa _{(t)}, \Omega _{(t)}$ should be functions of the independent horizon parameters $ \tilde{a}_{\Delta}$ and 
 $ J_{ \Delta }$ only and following exactness condition should hold:
\begin{eqnarray} \label{exactness}
 \frac{ \partial \kappa _{(t)} }{ \partial J_{ \Delta }} = \frac{ \partial \Omega _{(t)} }{ \partial \tilde{a}_{ \Delta } }.
\end{eqnarray}
However, given any vector field, it is not guaranteed that these will be satisfied. In other words, not all vector fields are
 Hamiltonian. Vector fields $t^a$ for which these conditions are satisfied are admissible and there are infinite of them. We wish
 to find the class of admissible $t^a$ s by solving \eqref{exactness}. The essential point is to show the existence of a 
canonical live vector field. The horizon energy defined by this canonical live vector field is called the horizon mass. In order to 
proceed, we make the following change of variables for convenience:
$$ ( \tilde{a}_{\Delta},  J_{ \Delta }) \rightarrow (R_+ , R_-)$$
with
%\begin{eqnarray}
%\tilde{a} _{\Delta} &=& 2 \pi ( R_+ - \frac{R_-}{ \gamma }) \nonumber\\
%J_{ \Delta } &=& \frac{\pi}{l}\left[ 2 R_+ R_- - \frac{R_+ ^2 + R_-^2}{ \gamma } \right]
%\end{eqnarray}
%Or the inverse maps:
\begin{eqnarray}\label{map}
R_+ &=& \sqrt{ \frac{ \gamma l  }{ 2\pi \left( \gamma ^2 -1 \right) } \left( J_{ \Delta  } + \frac{ \gamma\,  \tilde{a}_{ \Delta }^2 }{8 \pi l} \right) } \nonumber\\
R_- &=& \gamma \sqrt{ \frac{ \gamma l  }{ 2\pi \left( \gamma ^2 -1 \right) } \left( J_{ \Delta  } + \frac{ \gamma\,  \tilde{a}_{ \Delta }^2 }{8 \pi l} \right) } - \frac{ \gamma \tilde{a} _{ \Delta }}{4\pi}
\end{eqnarray}
Now for $ \kappa _{ (t) } $ we wish to start with a sufficiently smooth function $ \kappa _0 $  of the horizon parameters. In 
general $ \kappa _{(\ell)} \neq \kappa _0 $. But we can always find a 
phase-function $\xi$ in $ t^a \triangleq \xi\ell^a- \Omega _{(t)} \varphi$ such that $ \kappa _{(\xi \ell )} = \kappa _0 $. Again, there 
is a canonical choice, supplied by the known solution, the BTZ one, in which there is a unique 
BTZ black-hole for each choice of the horizon parameters. We therefore set $ \kappa _0 = \kappa_{(t)}(\mathrm{BTZ})$, where $t^a$ is 
the global time translation Killing field of the BTZ space time, and express it in terms of the newly introduced coordinates:
$$ \kappa _0 =  \frac{R_+^2 - R_-^2}{R_+ l^2} $$
The angular velocity $ \Omega _{(t)} $ satisfying \eqref{exactness} comes out 
as $ \Omega _{(t)} = \dfrac{R_-}{lR_+}$. Using this value of angular velocity and equation \eqref{map}in \eqref{1stlaw} we have
\begin{eqnarray}\label{energy}
 \delta E ^{t}_{ \Delta } &=& \delta \,\left[ \frac{2\pi}{l^2}\, \left( R_+ ^2 + R_- ^2 - 2R_+  R_- / \gamma \right) \, \right]
\end{eqnarray}
Now, from equations \eqref{map} and \eqref{energy}, we have horizon mass in terms of the 
independent horizon parameters:
$$M_{ \Delta }\left(J_{\Delta}, \tilde{a}_{\Delta}\right) =  \dfrac{ \gamma J_{ \Delta}}{l}+ \dfrac{\gamma ^{2}\, \tilde a _{\Delta}^2}{8\pi l^2} - 
\dfrac{\tilde{a}_{\Delta}}{2l^2} \sqrt{\,l \gamma\,\left( \gamma ^2 -1\right) \left( J_{\Delta} + \dfrac{\gamma\, \tilde{a}_{\Delta}^2}{8\pi l}\right)}.$$
It is not difficult to check that this works for BTZ black hole. Restricting to BTZ values, 
this reads: $M_{ \Delta } = (M - J/ \gamma l\,)$. This exactly 
matches with the asymptotic charge $X^{(t)} _{ \infty}( \delta) = \delta\, \left( M - J/ \gamma l\right)$ associated 
with asymptotic time translation vector $t^a$ of BTZ space time as would have been expected. We must also note that the 
deformations of the conserved charges : angular momentum and mass under the influence of the parameter $ \gamma$ are 
exactly same as those stated in \cite{blagojevic_2003,blagojevic_2005,blagojevic_2008} and at 
the `chiral point' ($ \gamma =1$) angular momentum and the mass become proportional to each other with opposite sign. 

%%%%%%%%%%%%%%%%%%%%%%%%%%%%%%%%%%%%%%%%%%%%%%%%%%%%%%%%%%%%%%%%%%%%%%%%%%%%%%%%%%%%%%%%%%

\section{Covariant phase space realization of asymptotic symmetry algebra}\label{sec4}
It has been suggested that microscopic details which explain the thermodynamics of black holes is independent of any
 theory of quantum gravity. If this is taken seriously, it implies that the microstates that describe black hole spacetime 
can be understood from a principle which is expected to govern all quantum gravity theory. It then seems natural to use the 
arguments of symmetry. Whatever be the theory of quantum gravity, it must at least preserve a part of the symmetries of 
classical theory.  Study of asymptotic symmetries have been advocated to serve this purpose and has achieved 
striking success in reproducing the Bekenstein-Hawking formula. This issue was first addressed in the context 
of $2+1$ gravity (with negative cosmological constant) by \cite{strominger_1997}.

In this issue we note that diffeomorphisms which are gauges for any theory of gravity become
 physical symmetry at the boundaries of the space time manifold by physical requirements (boundary conditions). For example, in 
$3+1$ dimensional asymptotically flat space times one naturally identifies a time like vector field at 
asymptotic infinity as the unique time translation (Killing) as in Minkowski space time and fixes it once and for all. This 
fixes the diffeomorphisms partially and play the role of a physical symmetry. Only then we can associate a 
Hamiltonian or Noether charge with time which is the ADM mass. In \cite{brown_henneaux}, the authors considered 
diffeomorphisms generated by asymptotic vector fields which are a bit `relaxed Killing symmetries' of the asymptotic
 metric in a $2+1$ dimensional space time and showed that they form the pair of affine
 Witt algebra (2D conformal algebra, or deformation algebra of $S^1$) as opposed to $SO(2,2)$, the isometry
 group of AdS${}_3$. We will show that those vector fields actually generate flows in the phase space which 
are at least locally Hamiltonian and find the corresponding Hamiltonians (hence qualifying as physical symmetries), \emph{i.e.} 
charges in the covariant phase space framework. The preference for this frame work is firstly due to its manifest 
covariant nature and secondly for its immense calculational simplicity, as compared to canonical framework \cite{hota}.

According to the suggestion mentioned above, this immediately implies that the quantum theory describing 
the microstates of black holes is a conformal field theory. The simple use of central charges in the 
Cardy formula determines the asymptotic density of quantum states of black holes which have same
 mass and angular momentum and approach the asymptotic configuration of a classical BTZ black hole; and eventually 
the Bekenstein-Hawking result. We shall use the covariant phase-space formulation to compute black hole entropy in this theory.

Let us gather the essential details for the asymptotic analysis. For the BTZ solution \eqref{btz_sol}, the tetrads and connections
are given by:
$$e^0 = N\,dt , ~~ e^1=N^{-1}\,dr ~~\mbox{and}~~~ e^2 = r\,(\,d\phi + N_{\phi}\, dt\,)$$ and 
$$ A^0 = - N \,d \phi ,~~ A^1 = N^{-1}\, N_{\phi} \,dr ~~ \mbox{and}~~~A^2 = -\dfrac{r}{l^2}\,dt - r \,N_{\phi}\,d \phi , $$
where $N^2 = \left[\dfrac{r^2}{l^2} - \dfrac{M}{\pi}+ \dfrac{J^2}{4 \pi ^2 r^2}\right] , ~~ N_{\phi} = \left[\dfrac{J}{2 \pi r^2}\right]$ and the 
internal metric is $ \eta _{IJ} = \mathrm{diag} (+,-,-)$.\\
The asymptotic form of these variables match with the AdS ones as expected upto different orders of $1/r$ \cite{blagojevic_2003,blagojevic_2008}.
The asymptotic vector fields which generate diffeomorphisms preserving the asymptotic AdS structure (much milder than the BTZ solution) 
are given by: 
$$ \xi _n := \exp\, (in x_+)\, \left[ l \left( 1- \dfrac{l^2 n^2}{2 r^2}\right) \,\partial _{t} - inr \, \partial _r + 
\left( 1+ \dfrac{l^2 n^2}{2 r^2}\right) \,\partial _{\phi} \right]$$ with $n$ an integer and $ x_+ =( t/l + \phi\,)$. It is
easy to check that the vector fields satisfy the affine Witt algebra:
\begin{eqnarray}\label{vir} 
\left[ \xi _ n, \xi _m\right] = -i (n-m)\, \xi _{n+m}
\end{eqnarray}
We now want to investigate if the algebra of the vector fields on the space-time manifold is also realised 
on the phase space \emph{i.e} the Hamiltonian functions (or the generators of diffeomorphisms) corresponding to the vector fields $ \xi^a_n$
also satisfy the affine algebra. To see this, we first associate a phase space vector field $\delta _{\xi _n}$ to each element $\xi _n$ of the algebra
such that $\delta _{\xi _n}$ acts 
as $ \pounds _{\xi _n}$ on dynamical variables\footnote{This is because vector fields on the space time manifold work as generators of 
infinitesimal diffeomorphisms}. Secondly, we need the symplectic structure which will enable
us to construct the Hamiltonian functions as has been described in the previous sections (see \eqref{sec3da} and \eqref{sec3e}). Since we are
interested in the asymptotic analysis, we will be interested in the contribution to the symplectic structure from the asymptopia or $S_{\infty}$.
If an internal boundary like NEH is present we can assume that the vector fields 
whose asymptotic forms are as $ \xi^a_n$ above vanish on that boundary. From this point of view, for any arbitrary vector field $\xi^{a}_{n}$ which
 vanish on any internal boundary (in this section, we shall reinstate $16\pi G$ but shall choose $c=h=1$):
\begin{equation}
8 \pi G \,\Omega ( \delta, \delta _{\xi }) =  \oint _{ S_{ \infty}}  \left[ \left( \xi \cdot e^I\right) \delta \underrightarrow{A_I}
 + \left( \xi \cdot A^I\right) \delta \underrightarrow{e_I}+\frac{l}{ \gamma}\left( \xi \cdot A^I\right) \delta \underrightarrow{A_I} 
+ \frac{1}{l \gamma} \left( \xi  \cdot e^I\right) \delta \underrightarrow{e_I}\right]
\end{equation}
The under right arrows indicate pull-back of the forms on $ S _{\infty}$. Therefore 
the second and the fourth term in the integral do not contribute. Only the 
internal component $e_2$ (as given above) survives under the pull back which is given by $ -r d\phi$. This being a 
phase space constant, the action of $\delta$ on it vanishes. Hence, we get
\begin{eqnarray} \label{charge}
8\pi G\,\Omega ( \delta, \delta _{\xi }) =  \oint _{ S_{ \infty}}  \left[ \xi \cdot (e^I+ \frac{l}{ \gamma} A^I)\right] \delta \underrightarrow{A_I}
\end{eqnarray}
for any arbitrary  vector field $ \xi $. Using the above expressions of the fields asymptotically, we have
$$8\pi G\,\Omega ( \delta, \delta _{\xi _n}) = \left(1- \frac{1}{ \gamma} \right)\delta \left(\,l\,M+J \,\right)\, \delta _{n,0}$$
hence $\delta_{ \xi _n}$ are at least locally hamiltonian for all $n$. 
We also note using \eqref{vir} that $\delta _{\left[ \xi _ n, \xi _m\right]}$ is also a Hamiltonian 
vector field with $\delta\,H[\{\xi_{n},\xi_{m}\}]$ given by the right hand side of the following equation
\begin{eqnarray}\label{vir2}
8\pi G\,\Omega ( \delta, \delta _{\left[ \xi _ n, \xi _m\right]}) = -i(n-m)\,\left(1- \frac{1}{ \gamma} \right)\delta \left(\,l\,M+J \,\right)
\,\delta _{m+n,0}
\end{eqnarray}
We shall now determine the current algebra of the 
Hamiltonian functions (\emph{i.e.} $\{H_{\xi_{n}},H_{\xi_{m}}\}$) generated by the Hamiltonian vector fields
$ \delta_{\xi _{n}}$ and $\delta_{\xi _{m}}$ for arbitrary $n,m$. This will be given by:
\begin{equation}
8 \pi G\,\Omega ( \delta _{\xi _m}, \delta _{\xi _n}) = \oint _{ S_{\infty} } \left[\xi _n \cdot (e^I +
 \frac{l}{ \gamma} A^I)\right]\,\delta _{ \xi _m} \underrightarrow{A_I}
\end{equation}
It is now important that we first pull back $ A _I$ and then calculate the action of $ \delta _{\xi _m}$ on it 
as Lie derivative. After some lines of calculation, we find:
\begin{eqnarray}\label{calc2}
8\pi G\,\Omega ( \delta _{\xi _m}, \delta _{\xi _n}) &=& - 2in \left(1- \frac{1}{ \gamma} \right)(\,J +lM\,)\, \delta _{m+n,0}
+ il \pi n^3 \,\left(1- \frac{1}{ \gamma} \right)\,\delta _{m+n,0} + \mathcal{O}\,(\frac{1}{r^2}) \nonumber\\
&=& - i(n-m) \left(1- \frac{1}{ \gamma} \right) (\,J +lM\,) \,\delta _{m+n,0} + il \pi n^3\,\left(1- \frac{1}{ \gamma} \right)\, \delta _{m+n,0}
\end{eqnarray}
Comparing \eqref{vir2} and \eqref{calc2} we infer that the asymptotic diffeomorphism algebra \eqref{vir} is exactly 
realized at the canonical level (as a current algebra) except a `central term' 
$  -il \pi n^3\left(1- \frac{1}{ \gamma} \right) \delta _{m+n,0}$. This is not surprising, although
 all the vector fields $\delta _{\xi _n}$ were Hamiltonian. The second cohomology group of the 
Witt algebra  \footnote{For any real lie algebra $\mathcal G$ and its dual $\mathcal G^*$ a skew symmetric 
bilinear map $ \alpha \in \mathcal G^* \wedge \mathcal G^*$ is said to be a {\it cocycle} 
if $\alpha \left(\left[A,B\right],C\right)+ \alpha \left(\left[B,C\right],A\right)+\alpha \left(\left[C,A\right],B\right)=0$ for 
all $A,B, C \in \mathcal{G}$ and $\left[,\right]$ is the usual product on $\mathcal G$. The
 elements $\eth f$ ($f \in \mathcal{G^*}$) defined via $ \eth f (A,B) = \frac{1}{2}f( \left[A,B\right])$, automatically 
cocycles by Jacobi identity, are called {\it coboundary}. Let us define an equivalence $\sim$ as: two
 cocycles $ \alpha \sim \beta$ if $ \alpha = \beta + \eth g$ for any $g \in \mathcal{G^*}$. Now one
 defines $H^2 \mathcal{G}$ as the additive group of equivalence classes found through the modulo action 
of the equivalence relation. All semi simple lie algebras have trivial second cohomology.} is not trivial. A theorem 
of symplectic geometry states that in this case the action of the algebra is not Hamiltonian and moment
maps donot exist, which on the other hand implies that the action of the lie algebra on phase space is not hamiltonian \cite{woodhouse} 
%\footnote{In the usual notation of Poisson brackets this reads $ \{h _m , h_n \} = h_{ \left[\xi _m , \xi _n \right]}$, where 
%$h_m$ are the hamiltonians with respect to $ \delta _{\xi _m}$},
\footnote{If $J[\xi_{m}]$ and $J[\xi_{n}]$ are Hamiltonians (calculated in the canonical phase-space) corresponding
to the vector fields $\xi_{m}$ and $\xi_{n}$,then $J[[\{\xi_{m},\xi_{n}\}]\neq \{J[\xi_{m}],J[\xi_{n}]\}$ 
where $\{J[\xi_{m}],J[\xi_{n}]\}=:\delta_{\xi_{n}}\,J[\xi_{m}]=-\delta_{\xi_{n}}\,J[\xi_{m}]$.}, \emph{i.e.}
\begin{eqnarray}\label{ham}
\delta \,\Omega \,(\, \delta _{ \xi _m}\, ,\, \delta _{ \xi _n}\,) \neq \Omega \,( \,\delta _{\,\left[ \xi _m , \xi _n\right]}\,,\, \delta\,) .
\end{eqnarray}

All of this calculation was done choosing the right moving vector fields.
There also are a set of left moving vector fields which preserve the asymptotic structure:
$$\tilde \xi _n := \exp (in x_-)\, \left[\, l \left( 1- \dfrac{l^2 n^2}{2 r^2}\right) \partial _{t} 
- inr \, \partial _r - \left( 1+ \dfrac{l^2 n^2}{2 r^2}\right) \,\partial _{\phi} \right]$$ where $ x _- = (t/l - \phi\,)$
Proceeding along the very same route as before, we again end up with the result that canonical realization of this asymptotic 
symmetries are also realized exactly upto a central term, which now becomes = $  -il \pi n^3\left(1+ \frac{1}{ \gamma} \right) \delta _{m+n,0}$

>From the definition of the central charge of Virasoro algebra, which is the centrally extended version of the
 Witt algebra, we arrive at the exact formulas for the central charges for the right and left moving algebras respectively : 
$$ c_{\pm} = \frac{3l}{2G}\left(1\pm \frac{1}{ \gamma} \right)$$
Once we have the central charges, we can apply the Cardy formula to the BTZ solution to obtain the black hole entropy:
\begin{eqnarray}\label{entropy}
S &=&\frac{2\pi\, r_{+}}{4G}-\frac{2\pi \,r_{-}}{4G\gamma}= \left( a_{\Delta}- \dfrac{l\pi J}{\gamma a_{\Delta}} \right)/4G \nonumber \\
  &=& \frac{ \tilde{a}}{4G} 
\end{eqnarray}
where $r_{+}$ and $r_{-}$ are the radii of the outer and inner horizon, respectively.
If we consider the thermodynamic analogy of the first law of black hole mechanics \eqref{1stlaw} (derived for general 
spacetimes only requiring presence of a weakly isolated horizon only from classical symplectic geometric considerations), we 
observe that $S \sim \tilde{a}$. Curiously, even in the quantum result \eqref{entropy}, the entropy-modified area relation continues to hold. 
%Also note that for $\gamma\rightarrow\infty$, we recover the usual result for $2+1$ gravity.
%%%%%%%%%%%%%%%%%%%%%%%%%%%%%%%%%%%%%%%%%%%%%%%%%%%%%%%%%%%%%%%%%%%%%%%%%%%%%%%%%%%%%%%%%%%%%%%%%%%%%%%%%%%%
\section{Conclusion}\label{sec5}

Let us recollect the main findings of this paper. Firstly, we introduced the concept of WIH in $2+1$ dimensions.
The boundary conditions which have been imposed on a $2$-dimensional null surface are much weaker than
the ones suggested in \cite{adw}. Our boundary conditions are satisfied by a equivalance class of null normals
which are related by functions, $[\xi\ell^{a}]$ rather than constants, $[c\ell^{a}]$ as was first proposed in \cite{adw}.
The advantage of such generalisation lies in the fact that it becomes possible to include extremal as well as non-extremal
solutions in the same space of solutions. Just by choosing the function $\xi$, one can move from a non-zero $\kappa_{(\ell)}$  
to a vanishing $\kappa_{(\xi\ell)}$ (see equation \eqref{kappa}) which essentially is like taking extremal limits 
in phase-space. We also established that the zeroth law (for all solutions in this extended space of solutions) follows quite trivially from the boundary conditions. 

Secondly, we have explicitly shown that in presence of 
an internal boundary satisfying the boundary conditions of a WIH, the variational principle for the generalised $2+1$ dimensional theory 
remains well-defined. This enable us to take the third step where we have constructed the covariant phase-space
of this theory. The covariant phase-space now contains all solutions of the $\gamma$-dependent theory 
which satisfy the WIH boundary conditions at infinity. As expected, extremal as well as non-extremal solutions form 
a part of this phase-space. We then went on to define the angular momentum as a Hamiltonian function corresponding
to the rotational Killing vector field on the horizon. It was also explicitly shown that for the BTZ solution,
the angular momentum defined in this manner matches with the expected result.

Thirdly, we established the first
law of black hole mechanics directly from the covariant phase-space, for isolated horizons. Instead of the usual horizon area term one encounters in this law, we find a modification due to the $ \gamma$ factor. This is a completely new result in this family of theories. It arose that the first law
is the necessarry and sufficient condition for existence of a timelike Hamiltonian vector field
on the covariant phase-space. However, not all timelike vector fields are Hamiltonian on phase-space,
there exists some which are admissible (there are in fact infinite of them). The canonical choice for 
these admissible vector fields are constructed too. Quite interestingly, the first law for the WIH formulation,
equation \eqref{1stlaw}, contains $\kappa_{(\xi\ell)}$. This implies that the first law
holds for all solutions, extremal as well as non-extremal. However, the thermodynamic implications of the first
law can only be extracted for non-extremal solutions since for the extremal ones, the first law is trivial. However,
we expect that since all solutions are equivalent from the point of view of WIH bounhdary conditions, the entropy of both class of black hole solutions will be same. 

Using asymptotic analysis, we have calculated the entropy of black holes for the theory under consideration. Contary to the usual approach, we construct 
the algebra of diffeomorphism generating Hamiltonian functions
directly from the covariant phase-space. As usual, we see that the algebra does not match with the Hamiltonian function for the
commutator of the asymptotic vector fields. The difference is the central extension. In other words, the algebra of spacetime vector fields
is not realised on the covariant phase-space. The Cardy formula then gives the entropy directly which matches
with the one expected from the first law. The entropy however not
only depends on the geometrical area but also on of other quantities like the parameter of the solution $J$ and the $\gamma$-parameter
of the theory (equation \eqref{entropy}). Keeping the thermodynamic analogy of laws of black hole mechanics in mind and concentrating on the BTZ black hole, one observes that there is a perfect harmony between this result and the modified first law. Also recall that our methods do not rely on existence of bifurcation spheres and applies equally to extremal and non-extremal black holes. To 
our knowledge, this has not been reproduced earlier since the phase space of Killing
horizons which satisfy laws of mechanics do not contain extremal solutions.        
 
Our analysis for the computation of entropy is based on asymptotic symmetry analysis. The principle
of using symmetry arguments to determine the density of states for black hole is attractive, it does
not depend on the details of quantum gravity. The asymptotic
analysis has a major drawback- it seems to be equally applicable for any massive object placed in place of
a black hole. Since such objects are not known to behave like black holes, it is not clear where to attribute
such large number of density of states. One must directly look at the near-horizon symmetry vector fields
for further understanding \cite{carlip_kh}. However, a more interesting step would be to determine the horizon microstates as is done in 
$3+1$ dimensions. In this case, it arises from classical considerations that the degrees of freedom that reside on a WIH in $3+1$ dimensions is
a Chern-Simons theory. Quantization of this theory gives an estimate of the states that contribute to a fixed area horizon and the entropy
turns out to be proportional to area. This has not been reproduced in $2+1$ dimensions still and will be investigated in future in order to compliment these new findings already present in this paper.

%%%%%%%%%%%%%%%%%%%%%%%%%%%%%%%%%%%%%%%%%%%%%%%%%%%%%
\section{Appendix}\label{ap1}
\subsection*{The Newman-Penrose formalism for $2+1$ dimensions}\label{New_Pen}
In order to make the article self-contained we summarise here the analogue of Newman-Penrose formalism in 2+1 dimensions,
 which was in detail described in \cite{adw}. We will use a triad
consisting of two null vectors $\l^a$ and $n^a$ and a {\it{real}}\footnote{All the N-P coefficients 
appearing in 2+1 dimensions are therefore real unlike in 3+1} space-like
vector $m^a$, subject to:
    \begin{eqnarray}\label{eqn:lmntetrad}
        \ell\cdot \ell = n\cdot n & = & 0,\ \ m\cdot m = 1\\
        \ell\cdot m & = & n \cdot m = 0\\
        \ell\cdot n & = & -1.
    \end{eqnarray}

The space-time metric $g_{ab}$ can be
expressed as
\begin{equation}\label{eqn:metric}
  g_{ab} = -2\, \ell_{(a}n_{b)} + m_am_b\, ,
\end{equation}
and its inverse $ g^{ab}$ is defined to satisfy
\begin{equation}\label{eqn:invmetric}
  g^{ab} = -2\, \ell^{(a}n^{b)} + m^am^b.
\end{equation}
It is then easy to verify that the expression for the triad is just
\begin{equation}\label{eqn:forme}
    e^I_a = -\ell_{a} n^I - n_a \ell^I + m_a m^I.
\end{equation}
Just as in the $3+1$ case, we express the connection in the chosen triad basis, the connection coefficients being the new N-P coefficients
(the $\gamma$ defined below is not to be confused with the Barbero-Immirzi parameter):
\begin{eqnarray}
    \nabla_a\ell_b & = & -\epsilon\, n_a\ell_b +\kappa_{\mbox{\scriptsize{NP}}}\, n_am_b -\gamma\,
    \ell_a\ell_b\nonumber \\
    & & \ \ + \tau\, \ell_a m_b + \alpha\, m_a\ell_b -\rho
    m_am_b\label{eq:nablal}\\
    \nabla_a n_b & = & \epsilon\, n_an_b -\pi\, n_am_b + \gamma\,
    \ell_an_b\nonumber\\
    & & \ \ - \nu\, \ell_am_b - \alpha\, m_an_b + \mu\, m_am_b\\
 \label{nablam}   \nabla_a m_b & = & \kappa_{\mbox{\scriptsize NP}}\, n_an_b - \pi\, n_a\ell_b + \tau\,
    \ell_a n_b\nonumber\\
    & & \ \  - \nu\, \ell_a \ell_b - \rho\, m_a n_b + \mu\, m_a \ell_b
\end{eqnarray}
It then simply follows from the expressions above that $
    \nabla_a\,\ell^a  = (\epsilon - \rho), \nabla_a\,n^a  = (\mu - \gamma) \mbox{and}
    \nabla_a\,m^a =  (\pi - \tau).$
Now we wish to expand the connection 1-form $A_a^I$ in the triad basis with 
N-P coefficients slated above as coefficients. In order to do so we note that for an arbitrary 1-form $v_a$ which 
may be mapped uniquely to an $SO(2,1) $ frame element $ v_I  = v_a e^a_I$. Then, for $
  \nabla_a v_b = {A_a^I}_J\, v^J\,  e_{Ib}$,
and using $ A_{a\ I}{}^{J} = \epsilon_{KI}{}^J A_a^K$, we arrive at the expression:
\begin{eqnarray}\label{eqn:anp}
    A_a^K  & = & (\pi n_a + \nu \ell_a - \mu m_a) \ell^K + (\kappa_{\mbox{\scriptsize NP}}\, n_a + \tau \ell_a - \rho m_a)\, n^K\nonumber\\
    &&+ (-\epsilon n_a - \gamma \ell_a +\alpha m_a ) \, m^K
\end{eqnarray}
%\subsection{Asymptotic boundary conditions}\label{Asymp} Ekhane Ashtekar et al b.c. thakbe
%\label{appb}
%jdsakljaos
\acknowledgements{RB thanks CSIR, India for financial support through the fellowship : SPM-07/575(0061)/2009-EMR-I and Samir K Paul for discussion.}  

%%%%%%%%%%%%%%%%%%%%%%%%%%%%%%%%%%%%%%%%%%%%%%%%%%%%%%%%%%%%%%%%%%%%%%%%%%%%

\end{document}